\documentstyle[12pt]{article}
\begin{document}

\begin{center}
{\bf Los Alamos Electronic ArXives} 

{\bf http://xxx.lanl.gov/physics/0005019\\}
%%%%%%%%%%%%%%%%%%%%%%%%%%%%%%%%%%%%%%%%%%%%%%%%%%%%%%%%%%%%%%%%%%%%%%%%%%%%%
%\\
%%%%%%%%%%%%%%%%%%%%%%%%%%%%%%%%%%%%%%%%%%%%%%%%%%%%%%%%%%%%%%%%%%%%%%%%%%%%%                                                                                                                                                                                  
                                                                                                  
\end{center}

\bigskip
\bigskip

\begin{center}
{\huge \bf Electrodin\'amica Cl\'asica}\end{center}

\bigskip

\begin{center}
{\large \bf HARET C. ROSU}\end{center}
\begin{center} e-mail: rosu@ifug3.ugto.mx\\
fax: 0052-47187611\\
phone: 0052-47183089  \end{center}
%Julio de 1998, April 2000}
%\end{center}

\bigskip
\bigskip

%\begin{center}
%{\LARGE {\bf Introducci\'on a la Electrodin\'amica Cl\'asica}}
%\end{center}

\bigskip
\bigskip
\bigskip

\begin{center}
{\bf Spanish Abstract}
\end{center}

\noindent
Aqu\'{\i} se bosquejan algunos temas impartidos durante el curso de 
electrodin\'amica cl\'asica I, de acuerdo al programa del Instituto
de F\'{\i}sica de la Universidad de Guanajuato, M\'exico.

\bigskip
\bigskip
\bigskip

\begin{center}
{\bf English Abstract}
\end{center}

\noindent
Excerpts are presented from a graduate course on Classical Electrodynamics 
held during the spring semester of 2000 at
the Institute of Physics, Guanajuato State University, Mexico.

\bigskip
\bigskip
\bigskip

\begin{center}
{\bf Mayo de 2000}
\end{center}

\begin{center} Copyright \copyright 2000 by the author.
%{\cal H}.{\cal C}. {\cal R}{\cal O}{\cal S}{\cal U}.
All commercial rights are reserved.
\end{center}

\newpage

\begin{center}
\section*{Clase 1}
\end{center}

\subsection*{Generalidades} 
Los campos el\'ectrico ($\vec{E}$) y de inducci\'on magn\'etica
($\vec{B}$) se introdujeron originalmente a trav\'es de la fuerza 
ejercida por cargas ($q'$) o corrientes ($I$) sobre una carga de prueba 
($q$):
\begin{eqnarray*}
\vec{F}_q = q\vec{E} &\leftrightarrow& \vec{E} = \frac{1}{q}\vec{F}_q,\\
d\vec{F}_I = Id\vec{l}\times\vec{B} &\leftrightarrow& B = 
\frac{|d\vec{F}_I|}{Idl\sin{\alpha}}.
\end{eqnarray*}
De acuerdo con esto, $\vec{E}$ se interpreta como {\it fuerza por unidad 
de carga} y $\vec{B}$ como {\it fuerza por unidad de corriente}. Sin 
embargo ambos campos tienen significado propio, independiente del tipo de 
fuente que los genera.

Ahora bien, $\vec{E}$ y $\vec{B}$ no son los \'unicos campos importantes 
en la electrodin\'amica. En la mayor{\'\i}a de las sustancias:
\begin{eqnarray*}
\vec{D} = \epsilon_0\vec{E} + \vec{P},\\
\vec{H} = \frac{1}{\mu_0}\vec{B} - \vec{M}~.
\end{eqnarray*}
$\vec{D}$ se conoce como desplazamiento el\'ectrico, $\vec{H}$ campo
magn\'etico, $\vec{P}$ y $\vec{M}$ son, respectivamente, las
polarizaciones el\'ectrica y magn\'etica (i. e., representan el promedio
macrosc\'opico de dipolos el\'ectricos$/$magn\'eticos en el material en
presencia de campos); $\epsilon_0 = 8.85\times10^{-12}C/Nm^2$, $\mu_0 =
4\pi\times10^{-7}H/m$ y $\epsilon_0\mu_0 = c^{-2}$. La conexi\'on entre
los vectores $(\vec{P}, \vec{E})$ y $(\vec{M}, \vec{H})$ est\'a
determinada por las propiedades de cada sustancia. Para medios
anisotr\'opicos la aproximaci\'on lineal en los campos es: 
\begin{eqnarray*}
P_i = \epsilon_0\alpha_{ik}E_k,\\
M_i = \kappa_{ik}H_k,
\end{eqnarray*}
con $i, k = 1, 2, 3$; $\alpha$ es el tensor de polarizabilidad y $\kappa$ 
el tensor de magnetizaci\'on. Entonces
\begin{eqnarray*}
D_i = \epsilon_{ik}E_k,\\
B_i = \mu_{ik}H_{k},
\end{eqnarray*}
donde $\epsilon_{ik} \equiv \epsilon_0(\delta_{ik} + \alpha_{ik})$, 
$\mu_{ik} \equiv \mu_0(\delta_{ik} + \kappa_{ik})$. Para medios 
isotr\'opicos:
\begin{eqnarray*}
&&\vec{P} = \epsilon_0\alpha\vec{E}\\
&&\vec{M} = \kappa\vec{H}, \quad \epsilon \equiv \epsilon_0(1 + \alpha)\\
&&\vec{D} = \epsilon\vec{E}, \quad \mu \equiv \mu_0(1 + \kappa)\\
&&\vec{B} = \mu\vec{H}.
\end{eqnarray*}

Una vez definidos estos vectores, podemos presentar las ecuaciones de 
Maxwell (1873), que son en electrodin\'amica lo que las leyes de Newton 
en mec\'anica. Las ecuaciones de Maxwell en forma diferencial son
\begin{eqnarray} 
&&\nabla\times\vec{H} = \vec{j} + \frac{\partial\vec{D}}{\partial{t}},
\label{ec1}\\
&&\nabla \cdot \vec{B} = 0,
\label{ec2}\\
&&\nabla\times\vec{E} = -\frac{\partial\vec{B}}{\partial{t}},
\label{ec3}\\
&&\nabla \cdot \vec{D} = \rho
\label{ec4}
\end{eqnarray}
($\rho$ es la densidad de carga y $\vec{j}$ la densidad de corriente). La 
forma integral de estas ecuaciones es
\begin{eqnarray*}
&&\oint_C\vec{H} \cdot d\vec{l} = \int_S\left(\vec{j} + 
\frac{\partial\vec{D}}{\partial{t}}\right) \cdot \hat{n}dA,\\
&&\oint_S\vec{B} \cdot \hat{n}dA = 0\\
&&\oint_C\vec{E} \cdot d\vec{l} = 
-\int_S\frac{\partial\vec{B}}{\partial{t}} \cdot \hat{n}dA,\\
&&\oint_S\vec{D} \cdot \hat{n}dA = \int_V\rho{dV}.
\end{eqnarray*}

De estas \'ultimas se obtienen las condiciones de frontera entre dos medios:
\begin{eqnarray}
&&(\vec{D}_2 - \vec{D}_1) \cdot \hat{n}_{1,2} = \sigma,
\label{ec5}\\
&&\hat{n}_{1,2}\times(\vec{E}_1 - \vec{E}_2) = 0,
\label{ec6}\\
&&(\vec{B}_2 - \vec{B}_1) \cdot \hat{n}_{1,2} = 0,
\label{ec7}\\
&&\hat{n}_{1,2}\times(\vec{H}_2 - \vec{H}_1) = \vec{i},
\label{ec8}
\end{eqnarray}
donde 
\[
|\vec{i}| = \left|\frac{dI}{dS}\right|.
\]

\subsection*{Electrost\'atica}
Estudiamos un problema de electrost\'atica si se satisfacen las condiciones
\begin{itemize}
\item No hay dependencia temporal en los campos.
\item No existen cargas en movimiento.
\end{itemize}
Con esto, las ecuaciones de Maxwell (\ref{ec1} - \ref{ec4}) se reducen a
\begin{eqnarray}
&&\nabla\times\vec{E} = 0,
\label{ec9}\\
&&\nabla \cdot \vec{D} = \rho.
\label{ec10}
\end{eqnarray}
En vista de (\ref{ec9}), del c\'alculo vectorial sabemos que
\[
\vec{E} = -\nabla\Phi.
\]
De esta forma se introduce el potencial electrost\'atico ($\Phi$). 
Considerando medios isotr\'opicos (i. e., $\vec{D} = \epsilon\vec{E}$) la 
ecuaci\'on (\ref{ec10}) se reduce a
\[
\nabla^2\Phi = -\frac{\rho}{\epsilon}
\]
que se conoce como {\it ecuaci\'on de Poisson} (en ausencia de cargas se 
obtiene la {\it ecuaci\'on de Laplace}).

Por otra parte, las condiciones de frontera (\ref{ec5} - \ref{ec8}) se 
reducen a 
\begin{eqnarray*}
\Phi_1 &=& \Phi_2,\\
\epsilon_1(\nabla\Phi\cdot\hat{n})_1 &-& 
\epsilon_2(\nabla\Phi\cdot\hat{n})_2 = \sigma.
\end{eqnarray*}

En el caso de un conductor, dado que en su interior el campo el\'ectrico 
es nulo, se tiene
\[
\Phi_{\mbox{{\small conductor}}} = \mbox{const},
\]
y as{\'\i}, la densidad superficial de carga en el mismo es
\[
\sigma = -\epsilon(\nabla\Phi\cdot\hat{n})_{\mbox{{\small afuera}}}.
\]

\subsection*{Magnetost\'atica}
Las condiciones para hablar de magnetost\'atica son: 
\begin{itemize}
\item No hay dependencia temporal en los campos.
\item Hasta hoy no se han detectado los monopolos magn\'eticos.
\end{itemize}
Bajo estas consideraciones las ecuaciones de Maxwell (\ref{ec1} - 
\ref{ec4}) se simplifican a
\begin{eqnarray}
&&\nabla\times\vec{H} = \vec{j},
\label{ec11}\\
&&\nabla \cdot \vec{B} = 0,
\label{ec12}
\end{eqnarray}
y las condiciones de frontera (\ref{ec5} - \ref{ec8})
\begin{eqnarray*}
&&(\vec{D}_1 - \vec{D}_2) \cdot \hat{n}_{1,2} = 0,\\
&&\hat{n}_{1,2}\times(\vec{H}_2 - \vec{H}_1) = \vec{i}.
\end{eqnarray*}

Al igual que en el caso electrost\'atico, a partir del c\'alculo 
vectorial y (\ref{ec12}) se introduce el potencial vectorial magn\'etico como
\[
\vec{B} = \nabla\times\vec{A}
\]
el cual, para materiales homog\'eneos e isotr\'opicos ($\vec{B} = 
\mu\vec{H}$), se obtiene de (\ref{ec11}) como
\begin{equation}
\nabla^2\vec{A} = -\mu\vec{j}
\label{ec13}
\end{equation}
(NOTA: cabe aclarar que, dada su definici\'on, da lo mismo tomar $\vec{A}$ 
que $\vec{A} + \nabla\varphi$; por ello se elige el potencial vectorial 
tal que $\nabla(\nabla \cdot \vec{A}) = 0$, obteniendo as{\'\i} 
(\ref{ec13}) a partir de (\ref{ec11})). Si se conoce $\vec{j}$, la 
soluci\'on a (\ref{ec13}) es
\[
\vec{A}(\vec{r}) = 
\frac{\mu}{4\pi}\int_V\frac{\vec{j}(\vec{r}')}{|\vec{r} - \vec{r}'|}dV'
\]
y para $r \gg r_{\mbox{{\small sistema}}}$
\[
\vec{A}(\vec{r}) = \frac{\mu}{4\pi}\frac{\vec{m}\times\vec{r}}{r^3}
\]
donde $\vec{m}$ es el momento magn\'etico del sistema, dado como
\[
\vec{m} = \frac{1}{2}\int_V\vec{r}\times\vec{j}(r)dV.
\]

Por \'ultimo, la energ{\'\i}a de un campo magn\'etico est\'atico es
\begin{eqnarray*}
W_{\mbox{{\small mag}}} &=& \frac{1}{2}\int\vec{B} \cdot \vec{H}dV\\
&=& \frac{\mu}{8\pi}\int\frac{\vec{j}(\vec{r}) \cdot 
\vec{j}(\vec{r}')}{|\vec{r} - \vec{r}'|}dVdV'.
\end{eqnarray*}

Para un sistema de conductores
\[
W_{\mbox{{\small mag}}} = \frac{1}{2}\sum_{i,k}L_{ik}I_iI_k,
\]
donde se define el coeficiente de inducci\'on magn\'etica entre las 
corrientes $\vec{j}_i$ y $\vec{j}_k$ como
\[
L_{ik} = \frac{\mu}{4\pi{I_i}I_k}\int\frac{\vec{j}_k(\vec{r}_k) \cdot 
\vec{j}_i(\vec{r}_i)}{|\vec{r}_k - \vec{r}_i|}dV_kdV_i~.
\]

%\end{document}

\newpage
%%%%%%%%%%%%%%%%%%%%%%%%%%%%%%%%%%%%%%%%%%%%%%%%%%%%%%%%%%%%%%%%%%%%%%%%%%%%%%

\newcommand{\ii}{\'{\i}}
\newcommand{\be}{\begin{equation}}
\newcommand{\ee}{\end{equation}}
\newcommand{\ol}{\overline}
\newcommand{\nab}{\bigtriangledown}

\begin{center}
\section*{Clase 2}
\end{center}
En el caso en el cual los campos var\'{\i}an lentamente en el tiempo, o sea
son funciones $f(at)$ donde se satisfacen las condiciones
\setcounter{equation}{0}
$$
a\ll 1\hspace{2cm}\omega\ll\frac{\sigma_c}{\epsilon}\hspace{2cm}l\ll\lambda
$$
con $\sigma_c$ conductividad, $\omega ,\lambda$ car\'acter\'{\i}sticas de las
oscilaciones electromagn\'eticas, $l$ dimensiones lineales del sistema.\\
Las ecuaciones de Maxwell toman la forma
$$
\nab\times{\bf H}={\bf J}\hspace{2cm}\nab\times{\bf E}=
-\frac{\partial{\bf B}}{\partial t}
$$
$$
\nab\cdot{\bf B}=0\hspace{2cm}\nab\cdot{\bf D}=\rho~.
$$
Obs\'ervese que se ha despreciado el  t\'ermino
$\frac{\partial{\bf D}}{\partial t}$.\\
En el caso de campos variables arbitrarios para situaciones en las cuales
no existen corrientes ni cargas presentes las ecuaciones de Maxwell toman
la siguiente forma

$$
\nab\times{\bf H}=\frac{\partial {\bf D}}{\partial t}\hspace{2cm}
\nab\times{\bf E}=-\frac{\partial{\bf B}}{\partial t}
$$
$$
\nab\cdot{\bf B}=0\hspace{2cm}\nab\cdot{\bf D}=0~.
$$
Se tienen soluciones tipo ondas planas
$$
{\bf E}={\bf E}_o  e^{i({\bf k}\cdot{\bf r}-\omega t)}
$$
$$
{\bf H}={\bf H}_o e^{i({\bf k}\cdot{\bf r}-\omega t)}~.
$$
Las notaciones usadas son las usuales, $\omega$ es la frecuencia,
$\mid{\bf k}\mid=\frac{\omega}{c}$ es el vector de onda, la direcci\'on
del cual, en medios isotr\'opicos, coincide con la direcci\'on de la
energ\ii a. El vector que justamente nos da el flujo de energ\ii a es el
llamado vector de Poynting ({\bf S}), definido por
\be
{\bf S}={\bf E}\times{\bf H}~.
\ee
Para campos variables la conexi\'on entre los campos y los potenciales es
de la forma
$$
{\bf E}=-\nab\phi-\frac{\partial{\bf A}}{\partial t}
$$
\be
{\bf B}=\nab\times{\bf A}~.
\ee
En general, los potenciales no son observables directamente (sino por sus
efectos, {\bf E},{\bf B}). Entre ellos existe una condici\'on muy importante
(de consistencia de la teor\ii a electromagn\'etica) que se llama condici\'on
de ``gauge" que puede ser diferente en funci\'on de la condici\'on 
considerada.\\
Una de las condiciones de ``gauge" m\'as frecuentes es la de Lorentz
\be
\nab\cdot{\bf A}+\epsilon\mu\frac{\partial\phi}{\partial t}=0~.
\ee
Esta condici\'on ``gauge" es muy usada porque permite una simple
generalizaci\'on de las ecuaciones laplacianas del caso est\'atico
$$
\Box\phi=-\frac{\rho}{\epsilon}
$$
\be
\Box{\bf A}=-\mu{\bf J}
\ee
estas ecuaciones son llamadas D'Alembertianas y
$\Box\equiv\nab^2-\frac{\partial^2}{\partial t^2}$.
Los potenciales que son soluci\'on de estas ecuaciones son llamados
potenciales retardados (y no precisamente porque sean muy tontos) los
cuales tienen la forma
\be
\phi ({\bf r},t)=\frac{1}{4\pi\epsilon}\int \frac{\rho ({\bf r}',t -
\frac{\mid{\bf r}-{\bf r}'\mid}{v})}{\mid{\bf r}-{\bf r}'\mid}dV'
\ee 
\be
{\bf A}({\bf r},t)=\frac{\mu}{4\pi}\int \frac{{\bf J}({\bf r}',t-
\frac{\mid{\bf r}-{\bf r}'\mid}{v})}{\mid{\bf r}-{\bf r}'\mid}dV'~.
\ee 
A grandes distancias del sistema de cargas ($r>>\lambda$) y en el vac\ii o,
{\bf B},{\bf E} y {\bf A} se pueden escribir como sigue
\be
{\bf B}=\frac{1}{c}\dot{{\bf A}}\times{\bf n}
\ee
\be
{\bf E}=c{\bf B}\times{\bf n}=(\dot{{\bf A}}\times{\bf n})\times{\bf n}
\ee
\be
{\bf A}=\frac{\mu}{4\pi r}\int {\bf J}({\bf r}',t-
\frac{\mid{\bf r}-{\bf r}'\mid}{v})dV'
\ee
donde {\bf n}$=\frac{{\bf r}}{r}$ es el versor en la direcci\'on de la
radiaci\'on. Si adem\'as $\lambda>>l$, con $l$ la dimensi\'on del sistema
radiante, se puede usar la llamada aproximaci\'on multipolar, es decir,
la radiaci\'on se puede representar como una sumatoria de los campos emitidos
por los dipolos, cuadrupolos, etc., que forman el sistema. Para el caso
dipolar se tiene
\be
{\bf B}=\mu_0\frac{\ddot{{\bf p}}\times{\bf n}}{4\pi c r}
\ee
\be
{\bf E}=\frac{\mu_0}{4 \pi r}(\ddot{{\bf p}}\times{\bf n}\times{\bf n})
\ee
donde {\bf p} es el momento dipolar del sistema.
La intensidad de la radiaci\'on de un dipolo es
\be
I=\frac{\ddot{{\bf p}}^2}{6\pi \epsilon_0 c^3}~.
\ee

\subsection*{Magnetohidrodin\'amica}

La magnetohidrodin\'amica estudia el comportamiento de los l\ii quidos o los 
gases conductores (plasmas) en campos electromagn\'eticos. Se usan los 
conceptos hidrodin\'amicos: densidad, velocidad, presi\'on, viscosidad. Las 
ecuaciones b\'asicas son:
$$
\frac{\partial\rho_m}{\partial t}+\nab\cdot(\rho_m{\bf v})=0
$$
$$
\rho_m\frac{\partial{\bf v}}{\partial t}+\rho_m 
({\bf v}\cdot\nab){\bf v}=-\nab P+{\bf j}\times{\bf B}
+\eta\nab^2{\bf v}+\rho_m {\bf g}
$$
$$
\nab\times{\bf E}=-\frac{\partial{\bf B}}{\partial t}\hspace{1cm}
\nab\times{\bf H}={\bf j}\hspace{1cm}{\bf j}=
\sigma_e({\bf E}+{\bf v}\times{\bf B})
$$
mas la ecuaci\'on de estado del fluido.

\subsection*{Relatividad Especial}

La teor\ii a de la relatividad especial surgi\'o en la electrodin\'amica
y se basa en dos postulados 
fundamentales
\begin{itemize}
\item
La velocidad de la luz en el vac\ii o $c=2.99793\times 10^8 m/s$ es una 
constante en todos los sistemas de referencia inerciales.
\item
Las leyes de la F\ii sica tienen la misma forma en todos los sistemas 
inerciales (covariancia de las leyes naturales).
\end{itemize}
Las transformaciones de Lorentz en una dimensi\'on se escriben as\ii
$$
x_1'=\frac{x_1+i\beta x_4}{\sqrt{1-\beta^2}}
$$
$$
x_2'=x_2\hspace{2cm}x_3'=x_3
$$
$$
x_4'=\frac{x_4-i\beta x_1}{\sqrt{1-\beta^2}}$$
donde $x_4=ict$, $x_4'=ict'$, $\beta=\frac{v}{c}$. Las velocidades $u'$ de un 
cuerpo en K' con respecto a las velocidades $u$ del mismo cuerpo en K est\'an 
relacionadas mediante las siguientes expresiones
$$
u_x'=\frac{u_x-v}{1-\frac{vu_x}{c^2}},\hspace{1cm}
u_y´=\frac{u_y\sqrt{1-\beta^2}}{1-\frac{vu_x}{c^2}},\hspace{1cm}
u_z'=\frac{u_z\sqrt{1-\beta^2}}{1-\frac{vu_x}{c^2}}~.
$$
La segunda ley para part\ii culas relativistas se escribe
$$
{\bf F}=\frac{d{\bf p}}{dt}=\frac{d}{dt}\left( \frac{m{\bf v}}{\sqrt{1-\beta^2}}
\right)~.
$$
Cantidades del tipo $({\bf p},i/cE)$ son llamadas cuadrivectores, el anterior 
se llama cuadrivector de impulso-energ\ii a. Otros ejemplos de cuadrivectores 
son $({\bf k},i/c\omega)$; $({\bf j},ic\rho)$; $({\bf A},i/c\varphi)$. Existen 
tambi\'en objetos llamados cuadritensores por extensi\'on de lo anterior, 
algunos ejemplos de ellos son
$$
F_{\alpha\beta}=
\left(
\begin{array}{rcll}
0&cB_z&-cB_y&-iE_x\\
-cB_z&0&cB_x&-iE_y\\
cB_y&-cB_x&0&-iE_z\\
iE_x&iE_y&iE_z&0
\end{array}
\right)
$$
$$
T_{\alpha\beta}=\Sigma_0(F_{\alpha\mu}F_{\beta\mu}-\frac{1}{4}
\delta_{\alpha\beta}F_{\mu\eta}F^{\mu\eta})~.
$$
%%%%%%%%%%%%%%%%%%%%%%%%%%%%%%%%%%%%%%%%%%%%%%%%%%%%%%%%%%%%%%%%%%%%%%%%%%%%%%%

\newpage

\begin{center}
\section*{Clase 3}
\end{center}

\bigskip

\subsection*{\bf Fuerza de Lorentz como fuerza lagrangiana}

\bigskip

Las ecuaciones del movimiento de Euler-Lagrange son 
\setcounter{equation}{0}
\begin{equation}
Q_{k}=-\frac{\partial L}{\partial q_{k}}+\frac{d}{dt}\left( \frac{\partial L%
}{\partial {\dot{q_{k}}}}\right) 
\end{equation}
donde las $Q_{k}$ son las fuerzas externas o fuerzas generalizadas y $L=T-U$ .

Por otra parte, las ecuaciones de Maxwell en unidades de Gauss son
\begin{eqnarray*}
\left( M1\right) \qquad\nabla \times \overrightarrow{E}+\frac{1}{c}%
\frac{\partial \overrightarrow{B}}{\partial t} &=&0 \qquad \qquad \qquad
\left( M3\right) \qquad \nabla \cdot 
\overrightarrow{D}=4\pi \rho  \\
\left(M 2\right) \qquad\nabla \times \overrightarrow{H}-\frac{1}{c}%
\frac{\partial \overrightarrow{D}}{\partial t} &=&\frac{4\pi }{c}%
\overrightarrow{j}\qquad \qquad \left( M4\right) \qquad
\nabla \cdot \overrightarrow{B}=0~.
\end{eqnarray*}

Ahora con $\overrightarrow{F}=q\overrightarrow{E}=-q\nabla \varphi $ s\'{o}%
lo en electrostatica en general la fuerza es la ley de Lorentz o sea
\begin{equation}
\overrightarrow{F}_{L}=q\left( \overrightarrow{E}+\frac{1}{c}\overrightarrow{%
v}\times \overrightarrow{B}\right)~. 
\end{equation}

Ahora de la $\left( M4\right) $ encontramos que $\overrightarrow{B}=\nabla
\times \overrightarrow{A}$ y sustituyendo en $\left( M1\right) $ encontramos 
\begin{equation}
\qquad \nabla \times \overrightarrow{E}+\frac{1}{c}\frac{\partial }{%
\partial t}\left( \nabla \times \overrightarrow{A}\right) =0\qquad
\end{equation}
por tanto
\begin{equation}
\nabla \times \left( \overrightarrow{E}+\frac{1}{c}\frac{\partial }{\partial
t}\overrightarrow{A}\right) =0
\end{equation}
de aqui que podemos definir una funci\'{o}n escalar tal que 
\begin{equation}
-\nabla \Phi =\overrightarrow{E}+\frac{1}{c}\frac{\partial }{\partial t}%
\overrightarrow{A}
\end{equation}
entonces
\begin{equation}
\overrightarrow{F}_{L}=q\left( -\nabla \Phi -\frac{1}{c}\frac{\partial }{%
\partial t}\overrightarrow{A}+\frac{1}{c}\overrightarrow{v}\times \left(
\nabla \times \overrightarrow{A}\right) \right) 
\end{equation}
donde el doble producto vectorial lo podemos expresar de la siguiente forma
\begin{equation}
\overrightarrow{v}\times \left( \nabla \times \overrightarrow{A}\right)
=\nabla \left( \overrightarrow{v}\cdot \overrightarrow{A}\right) -\frac{d%
\overrightarrow{A}}{dt}+\frac{\partial \overrightarrow{A}}{\partial t}
\end{equation}
por tanto 
\begin{eqnarray*}
\overrightarrow{F}_{L} &=&q\left( -\nabla \Phi -\frac{1}{c}\frac{\partial }{%
\partial t}\overrightarrow{A}+\frac{1}{c}\left( \nabla \left( 
\overrightarrow{v}\cdot \overrightarrow{A}\right) -\frac{d\overrightarrow{A}%
}{dt}+\frac{\partial \overrightarrow{A}}{\partial t}\right) \right)  \\
&=&q\left( -\nabla \left[ \Phi -\frac{1}{c}\overrightarrow{v}\cdot 
\overrightarrow{A}\right] -\frac{1}{c}\frac{d}{dt}\left[ \nabla _{%
\overrightarrow{v}}\left( \overrightarrow{v}\cdot \overrightarrow{A}\right) %
\right] \right) 
\end{eqnarray*}
lo que hace que $\overrightarrow{F}_{L}$ se pueda escribir como fuerza
lagrangiana
\begin{equation}
\overrightarrow{F}_{L}=-\nabla U+\frac{d}{dt}\frac{\partial U}{\partial 
\overrightarrow{v}}
\end{equation}
con $U=q\Phi -\frac{q}{c}\overrightarrow{v}\cdot \overrightarrow{A}$ .

\bigskip

\subsection*{Electrodin\'amica no l\'{\i}neal} 

\bigskip

Para la electrodinamica no l\'{\i}neal la constante dielectrica se
expresa como
\begin{equation} \label{**}
\varepsilon _{v}=\frac{\varepsilon _{o}}{\left( 1+\frac{1}{b^{2}}\left(
c^{2}B^{2}-E^{2}\right) ^{\frac{1}{2}}\right) }  %\tag{**}
\end{equation}
y la permeabilidad se escribe como
\begin{equation}
\mu _{v}=\mu _{o}\left( 1+\frac{1}{b^{2}}\left( c^{2}B^{2}-E^{2}\right) ^{%
\frac{1}{2}}\right) 
\end{equation}
donde $b$ en ambos casos es un parametro que fija una intensidad m\'{a}xima
de los campos.

Al menos para campos que var\'{\i}an lentamente, en funci\'{o}n de los
tensores de permeabilidades el\'{e}ctricas y magn\'{e}tica del vac\'{\i}o
tenemos
\begin{equation}
D_{i}=\sum_{k}\varepsilon _{ik}E_{k}{\rm \ \ \ \ y \ \ \ \ }%
B_{i}=\sum_{k}\mu _{ik}H_{k}
\end{equation}
donde 
\begin{eqnarray}
\varepsilon _{ik} &=&\varepsilon _{o}\left[ \delta _{ik}+\frac{e^{4}\hbar }{%
45\pi m^{4}c^{7}}2\left( E^{2}-c^{2}B^{2}\right) \delta
_{ik}+7c^{2}B_{i}B_{k}\right] +....  \\
\mu _{ik} &=&\mu _{o}\left[ \delta _{ik}+\frac{e^{4}\hbar }{45\pi m^{4}c^{7}}%
2\left( B^{2}-\frac{E^{2}}{c^{2}}\right) \delta _{ik}+7E_{i}E_{k}/c^{2}%
\right] +.....
\end{eqnarray}
para el l\'{\i}mite cl\'{a}sico hacemos $\hbar \rightarrow 0$ y estos efectos
no lineales desaparecen al comparar con la expresi\'{o}n cl\'{a}sica en (9) y
(10) encontramos
\begin{equation}
b_{q}=\frac{\sqrt{45\pi }}{2}\sqrt{\frac{e^{2}}{4\pi \varepsilon _{o}\hbar c}%
}\frac{e}{4\pi \varepsilon _{o}r_{o}^{2}}\approx 0.51\frac{e}{4\pi
\varepsilon _{o}r_{o}^{2}}=0.51\frac{e_{G}}{r_{o}^{2}}
\end{equation}
por tanto
\begin{equation}
r_{o}=\frac{e_{G}^{2}}{mc^{2}}\approx 2.8\times 10^{-15}\; {\rm metros}
\end{equation}
este es el radio cl\'{a}sico del electr\'{o}n.

%\ \  \textbf{Ley de Gauss\FRAME{ftbpF}{405.1875pt}{218.875pt}{0pt}{}{}{Figure 
%}{\special{language "Scientific Word";type "GRAPHIC";maintain-aspect-ratio
%TRUE;display "USEDEF";valid_file "T";width 405.1875pt;height 218.875pt;depth
%0pt;original-width 671.75pt;original-height 361.375pt;cropleft "0";croptop
%"1";cropright "1";cropbottom "0";tempfilename
%'FQMQEM01.wmf';tempfile-properties "XPR";}}}

Ahora si tenemos varias cargas
\begin{equation}
\oint \overrightarrow{E}\cdot \overrightarrow{n}\;da \quad=\frac{1}{%
\varepsilon _{o}}\sum_{i}q_{i}
\end{equation}
y si tenemos distribuciones de carga
\begin{equation}
\oint_{S}\overrightarrow{E}\cdot \overrightarrow{n}\;da\quad=\frac{1%
}{\varepsilon _{o}}\int_{V}\rho \left( \overrightarrow{x}\right) dV
\end{equation}
donde $V$ es el v\'{o}lumen enserrado por la superficie, ahora el teorema de
la divergencia nos dice que 
\begin{equation}
\oint_{S}\overrightarrow{v}\cdot \overrightarrow{n}\;da\quad%
=\int_{V}\nabla \cdot \overrightarrow{v}dV
\end{equation}
entonces aplicando este teorema en la ley de Gauss encontramos
\begin{equation}
\nabla \cdot \overrightarrow{E}=\frac{\rho }{\varepsilon _{o}}
\end{equation}
y esta es la forma diferencial de la ley de Gauss.

\newpage
%%%%%%%%%%%%%%%%%%%%%%%%%%%%%%%%%%%%%%%%%%%%%%%%%%%%%%%%%%%%%%%%%%%%%%%%%%%%%%

\begin{center}
\section*{Clase 5}
\end{center}

\subsection*{Energ{\'\i}a potencial electrost\'atica y densidad de 
energ{\'\i}a; capacitancia}
Imaginemos el caso en que una carga $q_i$ es tra{\'\i}da desde al 
infinito hasta el punto $\vec{x}_i$, localizado en una regi\'on del 
espacio donde se conoce el potencial electrost\'atico $\Phi(\vec{x})$. El 
trabajo realizado sobre esta carga es
\setcounter{equation}{0}
\[
W_i = q_i\Phi(\vec{x}_i).
\]
Ahora bien, si este potencial es provocado por la presencia de otras $n - 
1$ cargas, se tiene
\[
\Phi(\vec{x}_i) = \frac{1}{4\pi\epsilon_0}\sum^{n - 1}_{j = 
1}\frac{q_j}{|\vec{x}_i - \vec{x}_j|}
\]
y por tanto
\begin{equation}
W_i = \frac{q_i}{4\pi\epsilon_0}\sum^{n - 1}_{j = 1}\frac{q_j}{|\vec{x}_i 
- \vec{x}_j|}.
\label{ec1}
\end{equation}

Por un proceso mental similar, se puede ver que el trabajo total 
necesario para obtener el arreglo de $n$ cargas, trayendo cada una desde 
infinito a una regi\'on del espacio originalmente vac{\'\i}a, es
\begin{equation}
W_{\mbox{\small total}} = 
\frac{1}{8\pi\epsilon_0}\sum_i\sum_j\frac{q_iq_j}{|\vec{x}_i - \vec{x}_j|}
\label{ec2}
\end{equation}
donde $i, j$ toman todos los valores entre $1$ y $n$, excepto $i = j$ 
(autoenerg{\'\i}as).

En el caso de una distribuci\'on continua de cargas es claro que
\begin{equation}
W_{\mbox{\small total}} = 
\frac{1}{8\pi\epsilon_0}\int\int\frac{\rho(\vec{x})\rho(\vec{x}')}{|\vec{x} - \vec{x}'|}d^3xd^3x',
\label{ec3}
\end{equation}
expresi\'on que puede reescribirse de varias formas:
\begin{itemize}
\item En t\'erminos del potencial
\begin{equation}
W_{\mbox{\small total}} = \frac{1}{2}\int\rho(\vec{x})\Phi(\vec{x})d^3x.
\label{ec4}
\end{equation}
\item Utilizando la ecuaci\'on de Poisson:
\begin{equation}
W_{\mbox{\small total}} = -\frac{\epsilon_0}{2}\int\Phi\nabla^2\Phi{d^3x}.
\label{ec5}
\end{equation}
\end{itemize}

Integrando por partes la \'ultima expresi\'on se obtiene
\begin{eqnarray}
W_{\mbox{\small total}} &=& \frac{\epsilon_0}{2}\int|\nabla\Phi|^2d^3x 
\nonumber\\
&=& \frac{\epsilon_0}{2}\int|\vec{E}|^2d^3x.
\label{ec6}
\end{eqnarray}
Por la forma de la \'ultima integral, se define la densidad volum\'etrica 
de energ{\'\i}a como
\[
w = \frac{\epsilon_0}{2}|\vec{E}|^2.
\]

Notemos que esta densidad de energ{\'\i}a es no negativa, y por tanto el 
trabajo total tampoco ser\'a negativo. Sin embargo, de (\ref{ec1}) se ve 
que el trabajo para hacer un arreglo con dos cargas de signo contrario es 
negativo; esta contradicci\'on surge porque en las expresiones (\ref{ec3} 
- \ref{ec6}) se incluyen las autoenerg{\'\i}as en el trabajo total, 
mientras que en el caso discreto (\ref{ec2}) se las excluye.

Por \'ultimo, como siempre, se puede calcular la fuerza a partir de los 
cambios que sufre la energ{\'\i}a ante desplazamientos virtuales peque\~nos.

\bigskip

Consideremos un sistema de $n$ conductores, el $i-$\'esimo de ellos con 
carga $Q_i$ y potencial $V_i$. Dada la relaci\'on lineal que existe entre 
el potencial y la carga, podemos escribir
\[
V_i = \sum^{n}_{j = 1}p_{ij}Q_j,
\]
donde $p_{ij}$ depende s\'olo del arreglo geom\'etrico de los 
conductores. Invirtiendo las ecuaciones anteriores se obtiene
\[
Q_j = \sum^{n}_{i = 1}C_{ji}V_i.
\]
Los coeficientes $C_{ii}$ son las {\it capacitancias}, y $C_{ij}$ ($i 
\neq j$) los coeficientes de inducci\'on.

De esta forma
\begin{eqnarray*}
W_{\mbox{\small total}} &=& \frac{1}{2}\sum^{n}_{i = 1}Q_iV_i\\
&=& \frac{1}{2}\sum^{n}_{i, j = 1}C_{ij}V_iV_j.
\end{eqnarray*}

\subsection*{Aproximaci\'on variacional a la soluci\'on de las ecuaciones de 
Poisson y Laplace}
El uso de m\'etodos variacionales es muy popular en F{\'\i}sica. La 
electrodin\'amica no es la excepci\'on. En efecto, la idea de considerar 
funcionales cuyos extremales satisfagan ecuaciones de movimiento tipo 
Poisson o Laplace es muy sugestiva (sobre todo por la elegancia del 
m\'etodo variacional).

Consideremos la funcional
\begin{equation}
I[\psi] = \frac{1}{2}\int_V\nabla\psi \cdot \nabla\psi{d^3x} - 
\int_Vg\psi{d^3x},
\label{ec7}
\end{equation}
sujeta a la condici\'on tipo Dirichlet $\delta\psi(S) = 0$ ($S$ es la 
superficie cerrada que contiene a $V$). Es f\'acil ver que $\delta{I} = 
I[\psi + \delta\psi] - I[\psi] = 0$ conduce a la ecuaci\'on de movimiento
\[
\nabla^2\psi = -g.
\]
Se ve que este problema no es otro que resolver la ecuaci\'on de Poisson 
con condiciones de frontera tipo Dirichlet.

Similarmente, para condiciones de frontera tipo Neumann, se plantea el 
funcional
\begin{equation}
I[\psi] = \frac{1}{2}\int_V\nabla\psi \cdot \nabla\psi{d^3x} - 
\int_Vg{\psi}d^3x - \oint_Sf{\psi}d^3x,
\label{ec8}
\end{equation}
con
\[
\left(\frac{\partial\psi}{\partial{n}}\right)_S = f(S).
\]
Es f\'acil probar que $\delta{I[\psi]} = 0$ conduce a las ecuaciones
\begin{eqnarray*}
\nabla^2\psi = -g,\\
\left(\frac{\partial\psi}{\partial{n}}\right)_S = f(S).
\end{eqnarray*}

Resulta l\'ogico preguntar si este m\'etodo variacional de obtener la
ecuaci\'on de Poisson sirve para algo, o es s\'olo un juego matem\'atico.
Para contestar, notemos que una vez conocida la forma de los funcionales
(\ref{ec7}, \ref{ec8}) a\'un es necesario encontrar $\psi$ (o sea resolver
la ecuaci\'on de Poisson); por tanto el problema es el mismo. Sin embargo,
se pueden proponer soluciones $\psi = A\Psi(\vec{x}, \alpha, \beta, ...)$
que satisfagan las condiciones de frontera dadas, para despu\'es
variacionalmente encontrar las constantes indeterminadas (notar que con
esta elecci\'on $I = I[A, \alpha, \beta, ...]$). En este sentido el 
m\'etodo variacional sirve para encontrar soluciones aproximadas.

\newpage
%%%%%%%%%%%%%%%%%%%%%%%%%%%%%%%%%%%%%%%%%%%%%%%%%%%%%%%%%%%%%%%%%%%%%%%%%%%%%%%

\begin{center}
\section*{Clase 6}
\end{center}
\section*{M\'etodo de las im\'agenes}
Este m\'etodo se refiere a problemas de cargas puntuales en la presencia de 
superficies a potencial cero o constante. Las condiciones de frontera se 
simulan con cargas puntuales de valores y posiciones bien determinadas 
conocidas como ``cargas im\'agenes".\\

\subsection*{Carga puntual con esfera a $\phi =0$}	

El potencial asociado a la carga real y la carga imagen es
$$
\phi ({\bf x})=\frac{1}{4\pi\epsilon_0}[\frac{q}{\mid {\bf x}-
{\bf y}\mid}+\frac{q'}{\mid {\bf x}-{\bf y'}\mid}]~.
$$
La condici\'on de frontera es que el potencial se anule 
en $\mid{\bf x}\mid=a$. Introducimos dos vectores 
unitarios {\bf n},{\bf n'}, uno en la direcci\'on de {\bf x} y el 
otro en la direcci\'on de {\bf y}, de manera que el potencial se puede 
expresar
$$
\phi({\bf x})=\frac{1}{4\pi\epsilon_0}[\frac{q}{\mid x{\bf n}-
y{\bf n'}\mid}+\frac{q'}{\mid x{\bf n}-y´{\bf n'}\mid}]~.
$$
Factorizando $x$ del primer t\'ermino, $y'$ del segundo y valuando en $x=a$
$$
\phi(x=a)=\frac{1}{4\pi\epsilon_0}[\frac{q}{a\mid {\bf n}-\frac{y}{a}
{\bf n'}\mid}+\frac{q'}{y'\mid{\bf n'}-\frac{a}{y'}{\bf n}\mid}]~.
$$
Se observa que para que el potencial se anule en la frontera de la 
esfera se debe satisfacer
$$
\frac{q}{a}=-\frac{q'}{y'}~,\hspace{2cm}\frac{y}{a}=\frac{a}{y'}
$$
resolviendo estas ecuaciones se encuentra
$$
q'=-\frac{y'}{a}q=-\frac{a}{y}q,\hspace{2cm}y'=\frac{a^2}{y}
$$
$q'$ es la carga total de inducci\'on sobre la superficie de la esfera, 
podemos observar adem\'as lo siguiente
$$
y\rightarrow a \Rightarrow q'\rightarrow -q
$$
$$
y\rightarrow \infty\Rightarrow q'\rightarrow 0
$$
La densidad superficial de carga est\'a dada por 
$$
\sigma =-\epsilon_0\frac{\partial\phi}{\partial x}\mid_{x=a}=\frac{q}{4\pi 
a^2}\frac{a}{y}\frac{1-\frac{a^2}{y^2}}{( 1+\frac{a^2}{y^2}-2\frac{a}{y}
\cos\gamma)^{3/2}}~.
$$
Es posible calcular tambi\'en la fuerza de atracci\'on hacia la esfera, 
la magnitud de la cual est\'a dada por 
$$
\mid{\bf F}\mid=\frac{1}{4\pi\epsilon_0}\frac{q^2}{a^2}\frac{a^3}{y^3}
\left ( 1-\frac{a^2}{y^2}\right)^{-2}~.
$$

\subsection*{Carga q en presencia de una esfera conductora cargada a Q, 
aislada}

El potencial para esta configuraci\'on se puede expresar as\ii
$$
\phi ({\bf x})=\frac{1}{4\pi \epsilon_0}\left[ \frac{q}{\mid {\bf x}-
{\bf y}\mid}-\frac{aq}{y\mid {\bf x}-\frac{a^2}{y^2}{\bf y}\mid}+
\frac{Q+\frac{a}{y}q}{\mid{\bf x}\mid}\right]~.
$$
La fuerza de atracci\'on en este caso es 
$$
{\bf F}({\bf y})=\frac{1}{4\pi\epsilon_0}\frac{q{\bf y}}{y^3}
\left[ Q-\frac{qa^3(2y^2-a^2)}{y(y^2-a^2)^2}\right]~.
$$

\subsection*{Carga q cerca de una esfera conductora a potencial constante}

Para la situaci\'on presente el potencial adopta la forma
$$
\phi ({\bf x})=\frac{1}{4\pi\epsilon_0}\left[ \frac{q}{\mid 
{\bf x}-{\bf y}\mid}-\frac{aq}{y\mid {\bf x}-\frac{a^2}{y^2}
{\bf y}\mid}\right]+\frac{Va}{\mid{\bf x}\mid}~.
$$
La fuerza de atracci\'on est\'a dada por 
$$
{\bf F}({\bf y})=\frac{q{\bf y}}{y^3}\left[ Va-\frac{1}{4\pi\epsilon_0}
\frac{qay^3}{(y^2-a^2)^2}\right]~.
$$

\subsection*{Esfera conductora en un campo el\'ectrico uniforme}

Un campo el\'ectrico uniforme es producido por ejemplo por dos 
cargas puntuales $\pm$Q localizadas en $z=\pm R$ para $R\rightarrow\infty$. Si 
ahora una esfera conductora es colocada en el origen, el potencial ser\'a el 
debido a las cargas $\pm$Q en $\mp$R y sus im\'agenes $\mp \frac{Qa}{R}$ en 
$z=\mp\frac{a^2}{R}$

$$
\phi =\frac{1}{4\pi\epsilon_0}\left[ \frac{Q}{(r^2+R^2+2rR\cos\theta)^{1/2}}
-\frac{Q}{(r^2+R^2-2rR\cos\theta)^{1/2}}\right]+
$$

$$
\frac{1}{4\pi\epsilon_0}\left[-\frac{aQ}{R\left(r^2+\frac{a^4}{R^2}+\frac
{2a^2r}{R}\cos\theta\right)^{1/2}}+\frac{aQ}{R\left( r^2+\frac{a^4}{R^2}-
\frac{2a^2r}{R}\cos\theta\right)^{1/2}}\right]~.
$$
Como $R>>r$ podemos desarrollar los denominadores
$$
\phi =\frac{1}{4\pi\epsilon_0}\left[ -\frac{2Q}{R^2}r\cos\theta+
\frac{2Q}{R^2}\frac{a^3}{r^2}\cos\theta\right]+\cdots
$$
Para $R\rightarrow\infty$, $\frac{2Q}{4\pi\epsilon_0 R^2}$ es el campo 
aplicado de manera que el potencial en este l\ii mite toma la forma
$$
\phi_{R\rightarrow\infty}=-E_0\left( r-\frac{a^3}{r^2}\right)\cos\theta=
-E_0z+\frac{a^3}{r^3}E_0 z~,
$$
donde el \'ultimo t\'ermino es el del "dipolo imagen". La densidad 
superficial de carga est\'a dada por 
$$
\sigma =-\epsilon_0 \frac{\partial\phi}{\partial r}\mid_{r=a}=3\epsilon_0 
E_0 \cos\theta~,
$$
la cual se anula al integrarla sobre la superficie
$$
\int \sigma da =0~.
$$

\subsection*{Funci\'on de Green para la esfera conductora}

Para problemas de Dirichlet con conductores $G({\bf x},
{\bf x'})/4\pi\epsilon_0$ puede ser interpretada como el 
potencial debido a la distribuci\'on superficial de carga inducida sobre 
la superficie por la presencia de una carga puntual (fuente) en 
el punto {\bf x'}. Por definici\'on la funci\'on de 
Green $G({\bf x},{\bf x'})$ satisface la ecuaci\'on
$$
\nab'^2G({\bf x},{\bf x'})=-4\pi\delta({\bf x}-{\bf x'})~.
$$
Para el caso de la esfera la funci\'on de Green est\'a dada por
$$
G_{esf}({\bf x},{\bf x'})=\frac{1}{\mid {\bf x}-{\bf x'}\mid}-\frac{a}
{x' \mid {\bf x}-\frac{a^2}{x'^2}{\bf x'}\mid}~.
$$
En coordenadas esf\'ericas lo anterior es
$$
G_{esf}({\bf x},{\bf x'})=\frac{1}{\left( x^2+x'^2-2xx'
\cos\gamma\right)^{1/2}}-\frac{1}{\left( \frac{x^2x'^2}{a^2}+a^2-2xx'
\cos\gamma\right)^{1/2}}~,
$$
$$
\frac{\partial G}{\partial n'}\mid_{x'=a}=-\frac{x^2-a^2}
{a \left( x^2+a^2-2ax\cos\gamma\right)^{3/2}}\sim \sigma~.
$$
Recordando la soluci\'on de la ecuaci\'on de Poisson con condiciones de 
Dirichlet para el potencial
$$
\phi ({\bf x})=\frac{1}{4\pi\epsilon_0}\int_V\rho ({\bf x'})G_D({\bf x},
{\bf x'})d^3x-\frac{1}{4\pi}\oint_S\phi({\bf x'})\frac{\partial G_D}
{\partial n'}da'
$$
usando esto, podemos escribir la soluci\'on general para el potencial de la 
esfera conductora para la cual conocemos el potencial en la frontera
$$
\phi_{esf}({\bf x})=\frac{1}{4\pi}\int\phi (a,\theta',\varphi')
\frac{a(x^2-a^2)}{\left(x^2+a^2-2ax\cos\gamma\right)^{3/2}}d\Omega' ~,
$$
donde $\cos\gamma=\cos\theta\cos\theta'+\sin\theta\sin\theta'
\cos(\varphi-\varphi')$. Para el interior de la 
esfera $x^2-a^2\rightarrow a^2-x^2$, y en el caso en el que se tienen 
distribuciones volum\'etricas de carga se tiene que tomar en cuenta la 
contribuci\'on de la integral de volumen.

\newpage
%%%%%%%%%%%%%%%%%%%%%%%%%%%%%%%%%%%%%%%%%%%%%%%%%%%%%%%%%%%%%%%%%%%%%%%%%%%%%

\begin{center}
\section*{Clase 9}
\end{center}

\begin{center}
\subsection*{An\'alisis de elemento finito para resolver la ecuaci\'on de 
Poisson}
\end{center}

A continuaci\'on presentamos una breve introducci\'on al an\'alisis de 
elemento finito para resolver la ecuaci\'on de Poisson. Por simplicidad en 
la presentaci\'on s\'olo consideramos problemas bidimensionales.

Primeramente esbozamos el m\'etodo de Galerkin para replantear la
ecua-ci\'on de Poisson, y dividir la regi\'on de estudio en una red cuyo
n\'umero de celdas es finito. Por \'ultimo presentamos dos tipos
particulares de redes: cuadriculada regular y triangular. 

\bigskip

\subsection*{El m\'etodo de Galerkin}
Sea una regi\'on bidimensional $R$ limitada por una curva cerrada $C$; 
consideremos en $R$ la ecuaci\'on de Poisson
\setcounter{equation}{0}
\begin{equation}
\nabla^2\psi = -g
\label{ec1}
\end{equation}
con condiciones de frontera tipo Dirichlet; multiplicamos (\ref{ec1}) por 
una funci\'on de prueba $\phi(x, y)$ que sea continua a trozos en $R$ y 
tal que $\phi(C) = 0$; despu\'es integramos sobre $R$, obteniendo
\[
\int_R[\phi\nabla^2\psi + g\phi]dxdy = 0.
\]
A continuci\'on, utilizando la primera identidad de Green 
(bidimensional), la integral anterior se reescribe como
\begin{equation}
\int_R[\nabla\phi \cdot \nabla\psi - g\phi]dxdy = 0.
\label{ec2}
\end{equation}

El siguiente paso es dividir la regi\'on $R$ por medio de una red con $N$
celdas, y definir un conjunto de funciones $\{\phi_i(x, y), i = 1, 2, ...,
N\}$ tal que cada una de ellas es no nula s\'olo en una celda particular 
de la red. A continuaci\'on se expresa $\psi$ como
\[
\psi(x, y) \approx \sum_{i = 1}^N\Psi_i\phi_i(x, y);
\]
sustituyendo lo anterior en (\ref{ec2}) y escogiendo $\phi = \phi_j$ se 
obtiene
\[
\sum_{i = 1}^N\Psi_i\int\nabla\phi_i(x, y) \cdot \nabla\phi_j(x, y) = 
g_0\int_R\phi_i(x,y)dxdy~,
\]
donde se ha supuesto que las celdas son suficientemente peque\~nas como 
para que $g(x, y) \approx g_0$ dentro de ellas (el valor de $g_0$ 
var{\'\i}a de celda a celda). Con esto, (\ref{ec2}) se reduce a la 
ecuaci\'on matricial
\begin{equation}
{\bf K}\Psi = G
\label{ec3}
\end{equation}
aqu{\'\i} ${\bf K}$ es una matriz $N \times N$ con elementos
\[
k_{ij} \equiv \int_R\nabla\phi_i \cdot \nabla_j{dxdy}
\]
$\Psi$ es la matriz columna formada con los coeficientes $\Psi_i$; $G$ es 
una matriz columna con elementos
\[
G_i \equiv g_i\int_R\phi_i(x, y)dxdy.
\]
El poder del m\'etodo de Galerkin radica en que, por la forma como se 
escojen las $\phi_i$, la matriz ${\bf K}$ es dispersa, i.e., s\'olo 
pocos de sus elementos son diferentes de cero, y por ello es 
relativamente f\'acil conocer $\Psi$ a partir de (\ref{ec3}), lo cual nos 
da la soluci\'on a la ecuaci\'on tipo Poisson (\ref{ec1}).

\subsection*{Casos particulares}
\subsection*{Red cuadriculada regular}
Se escoge una red de cuadros, cada uno de lado $h$; sean $(x_i, y_j)$ las
coordenadas de los v\'ertices. Se toman las funciones $\phi_{ij}(x, y)$
tales que $\phi_{ij} \neq 0$ s\'olo en una vecindad de \'area $h^2$
alrededor de $(x_i, y_j)$, y las $\phi_{ij}$'s son linealmente
independientes entre s{\'\i}.  Con esto, de acuerdo al m\'etodo de
Galerkin
\[
\psi \approx \sum_{k, l = 1}^{(N_0)}\Psi_{kl}\phi_{kl}(x, y)
\]
donde se supone que el total de celdas es $N_0$; los coeficientes 
$\Psi_{kl}$ se obtienen a partir de (\ref{ec3}) con
\begin{eqnarray*}
&&{\bf K} = \left(\int_R\nabla\phi_{ij} \cdot \nabla\phi_{kl}dxdy\right),\\
&&(G) = \left(g(x_i, y_j)\int_R\phi_{ij}dxdy\right)\\
&&(\Psi) = (\Psi_i).
\end{eqnarray*}

La inconveniencia del uso de redes como \'esta es que se presentan casos 
donde el potencial var{\'\i}a de formas diferentes en diferentes 
regiones, y por ello ser{\'\i}a m\'as conveniente utilizar celdas 
irregulares. A continuaci\'on se presenta una de ellas.

\subsection*{Red triangular}
Las redes triangulares son las m\'as utilizadas en el an\'alisis de
elemento finito, por las razones expuestas al final de la secci\'on
anterior. Para este tipo de redes se asume que el elemento triangular
($e$) es lo suficientemente peque\~no como para que $\psi$ cambie poco en
su interior y de hecho pueda ser aproximado de forma lineal en cada
direcci\'on: 
\[
\psi(x, y) \approx \psi_e(x, y) = A + Bx + Cy.
\]
Sean $(x_i, y_i)$ ($i = 1, 2, 3$) las coordenadas de cada v\'ertice del
tri\'angulo. Entonces las constantes $(A, B, C)$ quedan determinadas por los
valores de $\psi$ en cada uno de ellos. 

Con el fin de sistematizar el procedimiento, es conveniente definir las
{\it funciones de forma} $N_j(x, y)$ (una por cada v\'ertice), tales que
$N^{(e)}_j(x_j, y_j) = 1$, $N^{(e \neq e_j)}_j(x, y) = 0$ y $N^{(e)}_j(x,
y) = 0$ si $x \neq x_j$, $y \neq y_j$. Por la linealidad de $\psi$ dentro
de $e$, tomamos $N^{(e)}_j(x, y) = a_j + b_jx + c_jy$. De aqu{\'\i}, 
para $j = 1$
\begin{eqnarray}
&&a_1 + b_1x_1 + c_1y_1 = 1 \nonumber\\
&&a_1 + b_1x_2 + c_1y_2 = 0\\
&&a_1 + b_1x_3 + c_1y_3 = 0 \nonumber
\label{ec5}
\end{eqnarray}
de donde
\begin{eqnarray*}
&&a_1 = \frac{1}{2S_e}(x_2y_3 - x_3y_2)\\
&&b_1 = \frac{1}{2S_e}(y_2 - y_3)\\
&&c_1 = \frac{-1}{2S_e}(x_2 - x_3)
\end{eqnarray*}
donde $S_e$ es el \'area del tri\'angulo $e$.

Ahora, siguiendo el m\'etodo de Galerkin, tomamos $\phi_i = N^{(e)}_i$. 
De esta forma, expresamos $\psi$ como
\begin{equation}
\psi(x, y) \approx \sum_{f, j}\Psi^{(f)}_jN^{(f)}_j(x, y),
\label{ec6}
\end{equation}
donde la suma se realiza para todos los tri\'angulos ($f$) y todos los 
v\'ertices de cada tri\'angulo ($j$); $\Psi^{(f)}_j$ es el valor de 
$\psi$ en el v\'ertice $j$ del tri\'angulo $f$. Estos coeficientes se 
encuentran, para cada tri\'angulo, a partir de una ecuaci\'on similar a 
(\ref{ec3}):
\[
\sum_{j = 1}^3k_{ij}^{(e)}\Psi_j^{(e)} = \frac{1}{3}S_eg_e
\]

\noindent
con $k^{(e)}_{ij} \equiv S_e(b_ib_j + c_ic_j)$ (coeficentes de
acoplamiento); $g_e \equiv g(\bar{x}_e, \bar{y}_e)$, y $(\bar{x}_e,
\bar{y}_e)$ son las coordenadas del centro de gravedad del tri\'angulo. A
continuaci\'on s\'olo falta incluir todos los tri\'angulos de la red. Para
ello, considerando que los v\'ertices interiores a $C$ son $N$, y el total
de v\'ertices (interiores a $C$ y sobre ella) es $N_0$, los {\'\i}ndices
corren de $1$ a $N$ para los v\'ertices internos, y de $N + 1$ a $N_0$
para los que est\'an sobre la frontera. Con esto, se obtiene la ecuaci\'on 
equivalente a (\ref{ec3}) para toda la red es con
\begin{eqnarray*} 
&&{\bf K} = (k_{ij}), \quad k_{ii} = \sum_Tk_{ii}^{(e)}, \quad k_{ij} = 
\sum_Ek_{ij}, i \neq j,\\
&&G_i = \frac{1}{3}\sum_TS_eg_e - \sum_{j = N + 
1}^{N_0}k_{ij}^{(e)}\Psi_j^{(e)};
\end{eqnarray*}
$T$ indica que la suma es sobre los tri\'angulos con v\'ertice com\'un 
$i$; $E$ que la suma es sobre tri\'angulos con lados entre los v\'ertices 
$i$, $j$.

Como ya se dijo, ${\bf K}$ es una matriz dispersa, y por tanto la
soluci\'on a (\ref{ec1}) se puede encontrar como
\[
\psi(x, y) \approx \sum_{f, j}\Psi^{(f)}_jN^{(f)}_j(x, y).
\]

\end{document}